\documentstyle[amssymb,aps,multicol,prl,epsf]{revtex}

\begin{document}
\title{``Cold spots'': a new model for transport in high $T_c$ cuprates}
\author{L. B. Ioffe$^{1}$ and A. J. Millis$^{2}$}
\address{$^{1}$ Department of Physics and Astronomy \\
Rutgers University, Piscataway, NJ 08854\\
$^2$ Department of Physics and Astronomy, The Johns Hopkins University\\
3400 North Charles St., Baltimore, MD 21218}
\maketitle

\begin{abstract}
We present a Boltzmann equation analysis of the transport properties of a
model of electrons with a lifetime which is short everywhere except near the
Brillouin zone diagonals. The anomalous lifetime is directly implied by
photoemission and c-axis transport data. We find quantitative agreement
between calculations and ac and dc longitudinal and Hall resistivity, but
the predicted longitudinal magnetoresistance disagrees with experiment. A
possible microscopic origin of the anomalous lifetime is discussed.
\end{abstract}


\begin{multicols}{2}

\section{Introduction}

Normal state transport in high-$T_{c}$ superconductors remains a
controversial subject. There are two broad classes of theoretical approaches:
the (generalized) fermi liquid, in which the basic current carrying 
entities are electrons, and the non fermi liquid in which the basic entities
are more exotic objects, e.g. spinons and holons \cite{Baskaran87}, phase
fluctuations of a superconducting order parameter \cite{Doniach90,Emery95} or
fermions with definite charge conjugation symmetry \cite{Coleman96}.
In fermi-liquid based approaches one must invoke an anomalous scattering
mechanism. Various models have been proposed \cite
{Carrington92,Stojkovic96,Hlubina93}; a central concept is the ``hot spot'',
a small region on the fermi surface (typically taken to be near the $(\pi ,0)
$ and $(0,\pi )$ points) where the electron lifetime is unusually short and
has an anomalous temperature dependence. Hot spots arise e.g. in models
involving antiferromagnetic fluctuations strongly peaked at a particular
momentum transfer. Models involving hot spots have had some successes, but
have not led to complete and generally accepted descriptions of cuprate
physics \cite{Hlubina93,Ong97}.

In this paper we argue that the data are better described in terms of {\it %
cold spots}: small regions which we take to be near the zone diagonal and in
which we assume the lifetime is much {\it longer }than elsewhere on the
Fermi surface A somewhat similar model was recently put forward by
Zheleznyak et. al.\cite{Zheleznyak97} and the idea bears an intriguing
resemblance to the 'fermi segments' appearing in a phenomenological model of
preformed pairs \cite{Geshkenbein97} and found in recent $SU(2)$ gauge
theory calculations \cite{Lee97}

We do not at present have a controlled calculation from a microscopic model
which produces cold spots; we suggest however that they might arise from
strong $d_{x^{2}-y^{2}}$ pairing fluctuations of the type proposed in \cite
{Geshkenbein97}, and we present a leading order estimate along these lines
in section IV below. There are four phenomenological justifications. One is
the apparently successful description of transport and most aspects of
magnetotransport to be discussed at length below. A second justification
comes from photoemission. Experiments \cite{Shen95,Randeria97} on optimally
doped curprates show that for momenta parallel to $(\pi ,\pi )$ the electron
spectral function exhibits a reasonably well-defined quasiparticle peak,
suggesting relatively weak scattering. However, for momenta near $(\pi ,0)$
or $(0,\pi )$, there is no discernible peak (at $T>T_{c}$): the spectral
function is very broad, suggesting relatively strong scattering. Thus
photoemission implies a lifetime which is generically short but has a
pronounced angular dependence. A third justification comes from c-axis
transport. In optimally doped materials the observed $\rho _{c}$ is large,
and only weakly temperature dependent, while $\rho _{ab}$ is small, and
strongly temperature dependent. The anisotropy $\rho _{c}/\rho _{ab}$
increases strongly as T is decreased. Within a fermi liquid picture this is
explicable only if the two quantities are controlled by different parts of
the fermi surface. Band theory calculations have shown that this may occur
in high $T_{c}$ materials: the calculated between-planes hopping $t_{\perp }(%
\vec{p})$ is strongly momentum dependent, vanishing for momenta parallel to $%
(\pi ,\pi )$ and being maximal for momenta parallel to $(\pi ,0)$\cite
{Andersen94,Liechtenstein96}. Therefore the observation of very different
temperature dependences of $\rho _{ab}$ and $\rho _{c}$ implies that $\rho
_{ab}$ must be controlled by carriers with momentum nearly parallel to $(\pi
,\pi )$ and that these carriers must have a much longer lifetime than do
carriers on other parts of the Fermi surface. A fourth argument concerns the
anomalously small vortex viscosity measured by Parks {\em et al} \cite
{Parks95} or inferred from resistivity data \cite{Geshkenbein98}. This, it
has been argued, can only be explained by a strongly anisotropic scattering
rate.

The remainder of this paper is organized as follows. In section II we
present the model and derive an approximate solution. In section III we
compare the solution to data and present new experimental tests. In section
IV we discuss a possible microscopic origin. Section V is a conclusion.

\section{Model}

To motivate the model, we consider the c-axis conductivity $\sigma_c$. In
very anisotropic systems such as high-Tc superconductors an adequate
theoretical expression is

\begin{equation}
\sigma_c(T)=\int \frac{d^2p}{(2\pi)^2} t_{\perp}(p)^2
G_R(p,\omega)G_A(p,\omega) \frac{\partial f}{\partial \omega}
\label{sigma_c}
\end{equation}

Here $t_{\perp }(p)$ is the interplane hopping, $f(\omega )$ is the fermi
function and $G_{R,A}$ are the retarded and advanced in-plane Green
functions evaluated at the chemical potential, which in our conventions is $%
\omega =0$. Band structure calculations yield
\[
t_{\perp }(p)=\frac{t_{0}}{4}[\cos (p_{x})-\cos (p_{y})]^2
\]
with $t_{0}\approx 0.15\;eV$ for bilayer $BSCCO$. In a fermi liquid,

\begin{equation}
G_{R,A}(p,\omega) = Z/(\omega-\epsilon_p \pm i\Gamma_p) +G_{inc}  
\label{GFL}
\end{equation}

Here $1>Z>0$ is the quasiparticle renormalization factor, $\epsilon_p$ is
the (renormalized) fermi velocity and $\Gamma_p$ is a scattering rate. In a
conventional fermi liquid at low T, $\Gamma_p=(2\tau_{imp})^{-1}+\omega^2+T^2$
with $\tau_{imp}$ an impurity scattering time. Evaluating Eq. \ref{sigma_c}
using Eq. \ref{GFL} and neglecting the incoherent part yields

\begin{equation}
\sigma_c(T)=\int (dp) t_{\perp}(p)^2 N(p)Z^2(p) /2 \Gamma_p
\end{equation}

Here $(dp)$ denotes an integral along the fermi line. The main temperature
dependence in this formula comes from $\Gamma $, so in a conventional fermi
liquid in which quantities depend only weakly on position on the fermi
surface, $\rho _{c}\sim \Gamma (T)$ would have the same temperature
dependence as $\rho _{ab}=m\Gamma (T)/(ne^{2})$. To explain the observed
T-dependence of $\rho _{c}/\rho _{ab}$ one must either assume $Z\rightarrow 0
$ (i.e. non fermi liquid behavior) or invoke a strong angular dependence in $%
\Gamma $. In this paper we study the consequences of the latter assumption.
Because $\rho _{c}$ is large and has only a weak T-dependence in optimally
doped materials, over most of the Fermi surface $\Gamma $ must be large and
have only a weak temperature dependence. Because $\rho _{ab}$ is small and
temperature dependent $\Gamma $ must be small and temperature dependent in
the parts of the Fermi surface where $t_{\perp }(p)$ vanishes. Therefore we
assume that $\Gamma $ has a large, angular dependent part which vanishes
quadratically for momenta parallel to $(\pi ,\pi )$ and has negligible
frequency and temperature dependence. We parametrize the fermi surface by a
magnitude $p_{F}(\theta )$ and angle $\theta $, with $\theta =0$
corresponding to $\vec{p}$ along the zone diagonal and write
\begin{equation}
\Gamma (\theta ,T)=\frac{1}{4}\Gamma _{0}\sin ^{2}(2\theta )+\frac{1}{\tau {%
_{FL}}}  \label{gamma}
\end{equation}
Near the diagonals we approximate 
\begin{equation}
\Gamma (\theta ,T)=\Gamma _{0}\theta ^{2}+\frac{1}{\tau _{FL}}
\label{gammaapprox}
\end{equation}
The fermi-liquid scattering rate, $1/\tau _{FL}$, is taken to be the sum of
an impurity part and a $T^{2}$ part; explicitly 
\begin{equation}
\frac{1}{\tau _{FL}}=\frac{1}{\tau _{imp}}+\frac{T^{2}}{T_{0}}  \label{taufl}
\end{equation}
Here $T_{0}$ is an energy scale which in conventional Fermi liquid is of the
order of the Fermi temperature but 
which we take as an adjustable parameter;
we will find that it is small of the order of $15\;meV$ for optimally doped $%
YBaCuO$, suggesting that even the diagonal scattering is stronger than in a
usual Fermi liquid. Note that $\tau _{FL}$ may in pronciple have $\omega $
dependence as well but we assume that it is weak because  fluctuations
causing the scattering are slow.

The parameter $\Gamma _{0}$ may be estimated from photoemission experiments.
In optimally doped cuprates in the normal state one observes near the $%
(0,\pi )$ point a very broad weakly dispersing feature with an energy
half-width $>0.1eV$ (see e.g. Fig 6 of Ref \cite{Randeria97}), suggesting $%
\Gamma _{0}>0.4\;eV$. For our numerical estimates we take $\Gamma_{0}=0.6\;eV$
because it fits the optical data well.

To study the effects of this unusual scattering on the electrons we write a
relaxation-time Boltzmann equation for the quasiparticle distribution
function $f_p$.

\begin{equation}
[-i\omega +e(\vec{E}+\frac{\vec{v_p}}{c} \times \vec{B}) \cdot \partial_{%
\vec{p}}]f_p = -\Gamma (f_p-f^0)  \label{boltzmann}
\end{equation}

Because of the strong variation of $\Gamma$ we shall be mainly interested in
states near the zone diagonal. We therefore introduce coordinates along and
perpendicular to the diagonal; only the dependence perpendicular is
non-trivial. The Boltzmann equation becomes

\begin{equation}
[-i\omega+\Gamma(\theta,T)+\omega_c(\theta) \partial_{\theta}]\delta
f(\theta,\omega)=-e\vec{E}\cdot\vec{v}(\theta)\frac{\partial f} {\partial
\epsilon_p}  \label{boltztheta}
\end{equation}

This equation may be solved. Because it turns out that the physics is
dominated by $\theta$ near the zone diagonals we may approximate $%
\omega_c(\theta)$ and $v(\theta)$ by their values $v$ and $\omega_c$ at $%
\theta=0$. Then in the first quadrant and for $\vec{E}$ parallel to the zone
diagonal we have

\begin{equation}
\delta f(\theta ,\omega )=-\frac{eEv}{\omega _{c}}\frac{\partial f_{0}}{%
\partial \epsilon _{p}}\int_{-\infty }^{\theta }d\theta ^{\prime }\exp
\left[ -K(\theta ,\theta ^{\prime })\right]   \label{soltuion}
\end{equation}

with 
\begin{equation}
K(\theta,\theta^{\prime})=\frac{1}{\omega_c}[-i\omega+1/\tau_{FL}](\theta-%
\theta^{\prime})+ \frac{\Gamma_0}{3\omega_c}(\theta^3-\theta^{\prime 3})
\label{K}
\end{equation}

From the solution one may construct currents in the usual way. We have
chosen the electric field to be parallel to the zone diagonal so we have

\begin{equation}
j_{\parallel}=\frac{ep_F}{\pi^2}\int d\theta \delta f(\theta,\omega)
\label{jparallel}
\end{equation}
\begin{equation}
j_{\perp}=\frac{ep_F}{\pi^2}\int d\theta \theta \delta f(\theta,\omega)
\label{jperp}
\end{equation}

At $B=0$, $j_{\perp }=0$ and $j_{\parallel }=\sigma _{xx}E$ with 
\begin{eqnarray}
\sigma _{xx} &=&\frac{e^{2}v_{F}p_{F}}{\pi ^{2}}\int d\theta \frac{1}{%
-i\omega +1/\tau _{FL}+\Gamma _{0}\theta ^{2}}  \nonumber \\
&=&\frac{e^{2}v_{F}p_{F}}{\pi }\sqrt{\frac{\tau _{FL}}{\Gamma _{0}}}\frac{1}{%
\sqrt{1-i\omega \tau _{FL}}}  
\label{sigma_xx}
\end{eqnarray}

Eq. (\ref{sigma_xx}) predicts that if $\tau \sim T^{-2}$ then $\sigma \sim%
1/T$, as observed in optimally doped cuprates. Qualitatively this behavior
occurs because the conductivity is dominated by a small patch (width $T$) of
weakly scattered (lifetime $T^{-2}$) electrons. Similarly, the weak field
Hall conductivity is 
\begin{equation}
\sigma _{xy}=\frac{\sigma _{xx}}{4}\frac{\omega _{c}\tau _{FL}}{1-i\omega
\tau _{FL}}  \label{sigma_xy}
\end{equation}

and, expanding to ${\cal O}B^{2}$, 
\begin{equation}
\frac{\delta \rho _{xx}}{\rho }=\frac{5}{32}\omega _{c}^{2}\tau
_{FL}^{3}\Gamma _{0}+\frac{1}{16}(\omega _{c}\tau _{FL})^{2}
\label{delta_rho_xx}
\end{equation}

As we shall see in the next section, all of these formulas agree
quantitatively with experiment except for the magnetoresistance. If the
leading term in Eq (\ref{delta_rho_xx}) were absent, it would agree also.

These expressions apply in the weak field limit, in which an electron is
scattered many times before the magnetic field bends its orbit appreciably.
In the present context ''appreciably'' means that the electron leaves the
small patch of size p near the diagonal which dominates conduction, in
contrast to the usual case in which ''appreciably'' means ''completes of
order one cyclotron orbit.'' The criterion for the weak field limit $\delta
\rho /\rho \ll 1$ is

\begin{equation}
\omega _{c}\ll \omega _{c}^{*}=\frac{(1-i\omega \tau _{FL})^{3/2}}{\tau
_{FL}^{3/2}\Gamma _{0}^{1/2}}  \label{omega_c^*}
\end{equation}

In the high field, $\omega _{c}\gg \omega _{c}^{*}$ limit one may calculate
by dropping the linear terms in the argument of the exponential. Then
calculations give 
\begin{equation}
\sigma _{xx}=\frac{2.4e^{2}v_{F}p_{F}}{\pi ^{2}\sqrt{3}\omega _{c}}\left( 
\frac{3\omega _{c}}{\Gamma _{0}}\right) ^{2/3}  \label{sigma_xx_H}
\end{equation}
and 
\begin{equation}
\sigma _{xy}=\frac{e^{2}v_{F}p_{F}}{\sqrt{3}\pi \Gamma _{0}}
\label{sigma_xy_H}
\end{equation}

So $\rho _{xy}\sim \omega _{c}^{2/3}$. This result may be understood as
follows: one expects $\rho _{xy}\sim B/n_{eff}$; here the effective number
of carriers is the size of the conducting patch which in the high field
limit is of order $\theta \sim B^{-1/3}$.

\section{Application to data}

We now discuss the applicability of our results to data. We begin with
clean, optimally doped $YBa_{2}Cu_{3}O_{7}$. In one sample of this material
the resistivity at $T_{c}$ was about $50\;\mu \Omega cm$\cite{Orenstein90};
the mean interplane spacing of $5.5\AA $ implies $\sigma
_{xx}=10^{-3}\;\Omega ^{-1}$ per plane. Photoemission measurements \cite
{Shen95} yield a zone-diagonal velocity $v_{F}=1.3\;eV\AA $ and $p_{F}=0.6\;%
\AA ^{-1}$; use of these values and the observed $\sigma _{xx}$ per plane
implies $\sqrt{\Gamma _{0}/\tau _{FL}}\approx 60\;meV$. Our rough estimate $%
\Gamma _{0}\approx 0.6\;eV$ then implies $\tau _{FL}^{-1}\approx 6\;meV$ at $%
T=100\;K$ and therefore $T_{0}\approx 12\;meV$, although the uncertainties
are substantial. The observed linearity of the resistivity then implies $%
\tau _{FL}^{-1}\approx 24\;meV$ at $T=200\;K$. 
We next use Eq (\ref{sigma_xx}) to fit the observed frequency dependent
conductivity.  Calculation and data \cite{Orenstein90} are shown in Fig 1
for $T=200K$.  The best fit corresponds to $1/\tau_{FL} = 21\;meV$,
very close to the $24\;meV$ estimated from dc transport. We have also fit
$100\;K$ and $300\;K$ data; agreement is comparatively good and leads to a
$T^2$ dependence of $1/\tau_{FL}$.

\vspace{0.25cm}

\centerline{\epsfxsize=9cm \epsfbox{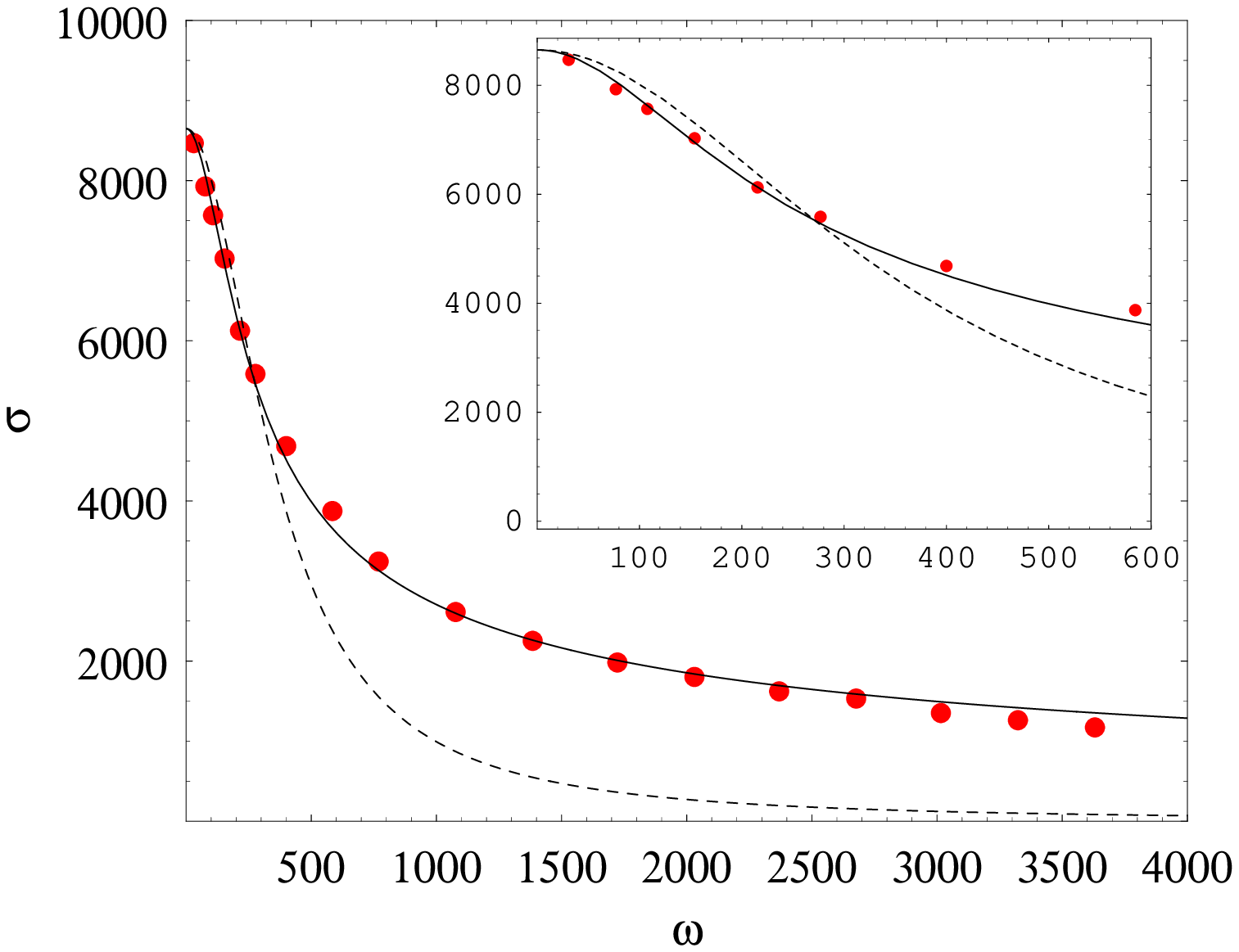}} 
{\footnotesize 
{\bf Fig. 1} Optical conductivity, $\sigma (\omega )$, at $T=200\;K$ in the 
units of $\Omega ^{-1}cm^{-1}$ plotted versus frequency ($cm^{-1}$): 
data from Ref \cite{Orenstein90} (shown as dots) 
and fit to Eq. (\ref{sigma_xx}) with
$\tau_{FL}^{-1}=21\;meV$ shown as solid line.  Drude fit with
$1/\tau =2T$ shown as dashed line.
}

\vspace{0.25cm}

The agreement between model and data
is very reasonable; we note especially the
narrowness of the Drude peak and the fact that upward concavity persists
down to a very low frequency. Many authors have argued that the observed
$\sigma(\omega)$ should be interpreted in terms of a frequency dependent
scattering rate $\tau_*^{-1}$ defined as 
\begin{equation}
\sigma(\omega) =
\frac{ne^2}{m^*(\omega)}\frac{1}{\tau_*^{-1}(\omega)-i\omega}
\label{sigma_exp}
\end{equation} 
In the dc limit $\tau_*^{-1}(\omega)$ is supposed to be proportional to $T$, with
constant of proportionality determined from the requirement that
$\frac{ne^2}{m^*}$ reproduce the London penetration depth. In the $YBa_2Cu_3O_7$
samples of \cite{Orenstein90} this implies $\tau_*^{-1} \approx 2T$. In the inset to
Fig. 1 we compare the data to the naive ansatz a frequency independent
scattering rate $\hbar/\tau = 2T$.
As can be this leads to seen in the inset, this leads to a conductivity which 
has too little upward concavity at low frequencies,
in contrast to Eq (\ref{sigma_xx}) which fits well over the entire frequency
range.
The observed upward concavity implies that the intrinsic low-frequency
scattering rate is much smaller at $T=200\;K$ than $\tau_*^{-1}=2T$ obtained
from the naive analysis. The discussion of $\sigma_2/\sigma_2$ to be presented
below will show that the intrinsic rate varies as $T^2$, not as $T$. Both of
these facts follow directly from our model but seem otherwise difficult to 
understand. 

We now turn to the high frequency behavior. The deviation between data and
naive Drude model visible in Fig. 1 has been previously been attributed to an
anomalous frequency dependent scattering although a generally accepted
microscopic derivation has not been found. One difficulty is that many
mechanisms for producing a $T$-linear resistivity involve scattering off
quasistatic fluctuations; such scattering leads to a too weak frequency
dependence of the scattering rate and therefore to a too rapid decrease of $%
\sigma_{1}(\omega )$. For example, in the gauge theory approach one finds $%
\sigma _{1}(\omega )\sim \frac{1}{T+\omega ^{\alpha }}$ with $\alpha =4/3$
or $3/2$ depending on whether spinons or holons dominate the transport
\cite{Ioffe90}.  Another difficulty is that in most models the $\omega$ and
$T$ dependences add, leading to a $T$ dependence of the high frequency
$\sigma$ which is inconsistent with the data.
By contrast, in the present approach the frequency dependence is essentially a
phase space effect and there is therefore only little $T$ dependence at large
$\omega$. An alternative model which gives the correct high $\omega$ behavior
is the Luttinger model of Anderson \cite{Anderson91}.

To further characterize the conductivity  we consider the
conductivity phase angle, $\frac{\sigma _{2}(\omega )}{\sigma _{1}(\omega )}$.
In Fig. 2 we show data for this quantity for $YBa_{2}Cu_{3}O_{7}$ at $%
T=95\;K$ that have been provided to us by Basov {\em et al}
and  data at $T=250\;K$ that we extracted from Ref. \cite{Collins89}.

Before proceeding with a more detailed analysis it should be noted that
Basov {\em at al} conductivity data were obtained on untwinned crystals with
the light polarized in the direction perpendicular to the chains and so this 
conductivity does not include contributions from the chains in contrast to 
the data of Ref \cite{Orenstein90,Collins89} obtained on twinned samples.
This is why the absolute magnitude of $\sigma _{1}$ at $\omega \lesssim
500\;cm^{-1}$  obtained by Basov {\em et al} are smaller in magnitude than
those reported by Orenstein {\em et al} by a factor of 2 at $T\approx
100\;K$. This makes a detailed comparison difficult. Qualitatively, 
however, we see that the phase angle tends to a value of the order of unity at
high frequencies, and that the crossover from the high frequency limit occurs
at a relatively low frequency. The fact that phase angle tends to a constant
implies that the real and imaginary parts of the conductivity vary as $%
\omega ^{-\alpha }$; that the constant is near unity imples that $\alpha
\approx 1/2$  (in general if $\sigma \propto \omega ^{-\alpha }$, $\sigma
_{2}/\sigma _{1}\rightarrow \tan (\alpha \pi /2)$) consistent with our form
Eq\ (\ref{sigma_xx}). We also note that high frequency limit  $\sigma \propto
1/\sqrt{\omega }$ is 
in a reasonable agreement with the empirical  observation \cite{Baraduc96}
that for many high-$T_{c}$ materials the high frequency conductivity obeys a
scaling $\sigma _{1}(\omega )\sim \omega ^{-\alpha }$ with $\alpha \approx
0.6$. We further note that the widespread practice \cite{Baraduc96} of
plotting $1/\tau ^{*}\equiv \frac{\omega \sigma _{1}(\omega )}{\sigma
_{2}(\omega )}$ is not very informative in such cases, the constant high
frequency limit of the ratio guarantees an apparently linear ''scattering
rate''. Despite the uncertainties
we have fit the observed frequency dependence of the phase angle to our
theoretically expected form $\tan \left[ \frac{1}{2}\tan ^{-1}(\omega \tau
)\right] $; the results are shown in Fig. 2 along with the data. 
Surprisingly,  $1/\tau_{FL}$ used in these fits agrees
well in magnitude with that determined from the dc conductivity or
the data of \cite{Orenstein90}. We emphasize that one sees directly from the
data in Fig. 2 that there is an intrinsic frequency scale in the problem which
varies more nearly as $T^2$ than as $T$ at low temperatures. This frequency
scale occurs naturally in our model.

\vspace{0.25cm}

\centerline{\epsfxsize=9cm \epsfbox{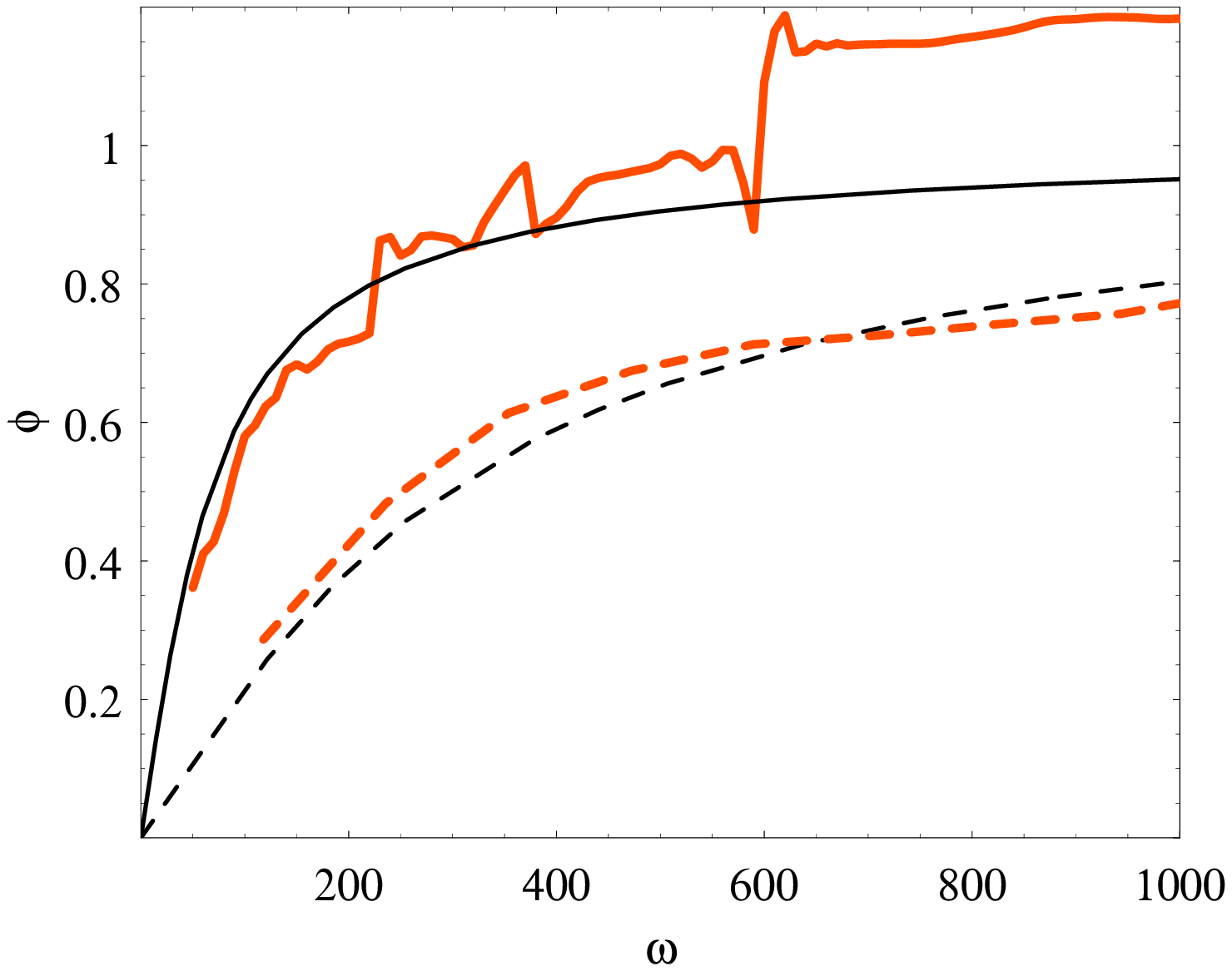}} 
{\footnotesize 
{\bf Fig. 2} Optical conductivity phase angle $\phi =\sigma _{2}/\sigma _{1}$
plotted versus frequency (cm$^{-1}$): data at $95\;K$ 
from \cite{Basov95} and at $250\;K$ from \cite{Collins89}
(thick dashed curve) and fit to Eq (\ref{sigma_xx}) with 
$\tau _{FL}^{-1}=6\;meV$, (thin curve) and $\tau _{FL}^{-1}=28\;meV$ (thin 
dashed curve).
}

\vspace{0.25cm}

We now turn to the weak-field Hall conductance. We have 
\begin{equation}
\cot (\Theta _{H})=\frac{1}{\omega _{c}\tau _{FL}}\left[ 1-i\omega \tau
_{FL}\right]   \label{cot_H}
\end{equation}
Thus the model yields the $\cot (\Theta _{H})\sim T^{2}$ law found
experimentally. More interestingly, it implies that over a wide frequency
range $Re\cot (\Theta _{H})\sim const$ and $Im\cot (\Theta _{H})\sim -\omega
/\omega _{c}$. Precisely this behavior has been observed experimentally (see
Fig. 4 of Ref. \cite{Kaplan96}). From these data we may infer that at $%
T=100\;K$, $B=8\;T$, $\omega _{c}\approx 1.25\;cm^{-1}$ and $\tau
_{FL}^{-1}\approx 6\;meV$, consistent with $\tau _{FL}^{-1}$ inferred from the
$\sigma _{xx}$ conductivity. Thus at $T=100\;K$ the scattering rate determined
from the ac Hall effect is numerically the same as that determined from the
zero field ac conductivity. Ac Hall data at other temperatures are not
available but the observed $T^2$ dependence of the dc Hall angle suggests to
us that all temperatures the ``Hall scattering rate'' is identical to our
$\tau_{FL}^{-1}$, and is not an independent quantity. Extension of the optical
Hall angle data to higher temperature and frequencies would be of great
interest: the present model predicts the frequency dependence 
should persist to high frequencies until the $\omega ^{2}$ term in $\tau
_{FL}^{-1}$ becomes important; a hint of this behavior may be discerned in
the data of Ref. \cite{Kaplan96} although the measured  $\tau ^{-1}(\omega )$
is constant within error bars.  Further, the only temperature dependence
should be a $T^{2}$ variation of $Re\cot (\Theta _{H})$.

An interesting question concerns the thermopower, which is
also anomalous in high-$T_c$ materials.  The anomalies
in the thermopower have been argued to scale in the same
way as the cotangent of the Hall angle \cite{Clayhold97}
 However, we do not have a reliable
calculation of thermopower in our model.

We turn now to the magnetoresistance, where a troubling discrepancy exists
between model and data. Assuming the magnetic field does not affect the
scattering mechanism we find 
\begin{equation}
\frac{\delta \rho _{xx}}{\rho _{xx}}=(2.5\Gamma _{0}\tau _{FL}+1)\tan
^{2}(\Theta _{H})  \label{delta_rho_reduced}
\end{equation}

Thus in this model the weak field magnetoresistance is very large compared
to the square of the tangent of the Hall angle and has an additional
temperature dependence. These results strongly contradict available data 
\cite{Harris95} which suggest that $\frac{\delta \rho }{\rho \tan
^{2}(\theta _{H})}$ is about $1$ at all temperatures in optimally doped and
underdoped $YBa_{2}Cu_{3}O_{7-\delta }$ and optimally doped and underdoped $%
Tl_{2}Ba_{2}CuO_{6+\delta }$. The observed ratio $\frac{\delta \rho }{\rho
\tan ^{2}(\theta _{H})}$ is much larger in $LaSrCuO_{4}$ but because this is
a very resistive material and we do not know how to interpret these data
within our model. We regard the $YBa_{2}Cu_{3}O_{7-\delta }$ and $%
Tl_{2}Ba_{2}CuO_{6+\delta }$ as more representative of the intrinsic
behavior of high $T_{c}$ materials.

There are two possible resolutions of this discrepancy. Either the model is
simply inapplicable, which seems unlikely in the view of the photoemission
results and the successes of the optics, or the anomalous scattering rate is
itself field dependent in such a way as to cancel the $\Gamma _{0}\tau $
term in Eq. (\ref{delta_rho_xx}). Because it is likely that the anomalous
scattering rate is related to pairing fluctuations a strong $B$ dependence
of the scattering rate is possible.

We briefly discuss impurity dependence. We expect $\tau _{FL}^{-1}=\tau
_{imp}^{-1}+T^{2}/T_{0}$ with the impurity scattering rate $\tau _{imp}^{-1}$
linearly dependent on the defect density $n_{d}$. Eq. (\ref{sigma_xx}) for $%
\sigma _{xx}$ then predicts strong violations of Matthiesson's rule: the $%
T=0 $ residual resistivity is predicted to go as the square root of the
defect density and the difference $\rho (T,n_{d1})-\rho (T,n_{d2})$ in
resistivity between two otherwise identical samples with different defect
density should be temperature dependent, decreasing as $T$ is increased.
Neither of these predictions is consistent with available data \cite
{Mizuhashi95}. We note, however, that the data mostly involve doping on the $%
Cu$ site. Such doping is known to drastically change the local environment
(e.g. by inducing magnetic moments on nearby $Cu$ sites \cite{Mendels94}).
Further, we note that in all available data, doping increases the slope $%
d\rho /dT$, presumably because the doping changes an effective carrier
number. These effects cannot be modelled by simply a $1/\tau _{imp}$ to a
theory; therefore we do not regard the inconsistency between the predicted
and observed doping dependence of $\rho $ as a definitive disproof of the
model we have proposed.

\section{Microscopic Origin}

In this section we present a qualitative discussion of a possible origin for
the unusual lifetime we proposed. The angular dependence is reminiscent of
that of the $d_{x^{2}-y^{2}}$ superconducting gap; we therefore propose that
the lifetime is caused by interaction of electrons with nearly singular $%
d_{x^{2}-y^{2}}$ pairing fluctuations. Similar ideas have been proposed by
Chubukov\cite{Chubukov97}. As an example of an interaction which yields the
desired behavior we consider 
\begin{equation}
\Gamma (p,p^{\prime },q)=\Gamma \sin 2\theta _{p}\sin 2\theta _{p^{\prime
}}D(\omega ,q)  \label{Gamma}
\end{equation}
with the pair fluctuation propagator $D(\omega ,q)$ given by 
\begin{equation}
D(\omega ,q)=\frac{\phi [\frac{\omega ^{2}}{u^{2}(q^{2}+\xi ^{-2})}]}{%
q^{2}+\xi ^{-2}}  \label{D}
\end{equation}

Here $\Gamma $ is an energy scale, $sin(2\theta _{p})$ is the usual d-wave
form factor and $\phi $ is a scaling function.  the form of $\phi$ is
not important; it is however
necessary to assume $\int dx \phi^{''}(x)$ is convergent and indeed of
order unity.   Note that we have assumed $%
\omega \sim uq$ scaling. This is a nontrivial (but widely made) assumption
for which we do not have a theoretical justification.

We may construct a one-loop self-energy from the interaction $\Gamma$ in the
usual way. We have (here $p$ and $q$ represent both momentum and frequency)

\begin{equation}
\Sigma(p) = \int (dq) \Gamma(p-q/2,p-q/2,q) G(p+q)  \label{sigma}
\end{equation}

We are interested in momenta near the zone diagonal where the self energy is
small. We assume $p_{F}\xi \gg 1$. We may therefore approximate $G(p+q)$ by
its weak coupling form $G(p)=(\omega -v_{F}(|p|-p_{F}))^{-1}$. After
integration over the magnitude of $|p+q|$ we obtain for the imaginary part $%
\Sigma ^{^{\prime \prime }}$ as a function of position $\theta $ on the
fermi line 
\begin{equation}
\Sigma ^{\prime \prime }(\theta )=\frac{p_{F}\Gamma }{v_{F}}\int \frac{%
d\theta ^{\prime }}{2\pi }\frac{d\omega }{2\pi }\sin ^{2}(2(\theta -\theta
^{\prime }))\frac{\phi ^{\prime \prime }(\frac{\omega ^{2}}{p_{F}^{2}\theta
^{\prime }{}^{2}+\xi ^{-2}})}{p_{F}^{2}\theta ^{^{\prime }}{}^{2}+\xi ^{-2}}
\label{sigma''}
\end{equation}

Now $\phi ^{\prime \prime }$ is peaked at $\omega \sim u\sqrt{(p_{F}\theta
)^{2}+\xi ^{-2}}$ and vanishes as $\omega \rightarrow 0$; thus the integral
is dominated by $\theta ^{\prime }\sim (p_{F}\xi )^{-1}$. for $\theta \gg
(p_{F}\xi )^{-1}$ we find $\Sigma ^{\prime \prime }(\theta )\sim \frac{%
u\Gamma }{v_{F}}\sin ^{2}(2\theta )$ while for $\theta \ll (p_{F}\xi )^{-1}$
we get $\Sigma ^{\prime \prime }(\theta )\sim \frac{u\Gamma }{v_{F}}%
(p_{F}\xi )^{-2}$. Further the $\omega \sim uq$ scaling implies $\xi \sim u/T
$. Thus the interaction leads to a scattering rate which is roughly of the
form $max(\Gamma _{0}\theta ^{2},T^{2}/T_{0})$ with $\Gamma _{0}\sim \frac{u%
}{v_{F}}\Gamma $ and $T_{0}\sim \frac{(up_{F})^{2}}{\Gamma _{0}}$ as
required for the previous analysis. The empirically determined $T_{0}\approx
12\;meV$ and $\Gamma _{0}\approx 0.6\;eV$ implies $u\sim 0.15\;eV\AA \sim
0.1v_{F}$. Therefore the one-loop approximation may be justified by the
usual Migdal arguments. Within the one-loop approximation the frequency
dependence of $\Sigma $ may be determined; it is very weak (logarithmic),
justifying its neglect in the phenomenological analysis.

The crucial features of the interaction are (i) it must be rather sharply
peaked about $q=0$ so as not to mix different parts of the Fermi surface too
strongly, (ii) it must be a singular function of $\omega $, so as to produce
a large, essentially $T$-independent scattering rate for $\theta $ away from
the zone diagonal and (iii) the basic scale (set by $\xi ^{-1}$) is $T$. We
expect these assumptions to apply for optimally doped materials. For
underdoped materials we conjecture that the system is closer to an
instability, i.e. $\xi ^{-1}\ll T$ implying a more singular interaction
which leads among other things to a pseudogap in the density of states (see 
\cite{Ioffe93} for a similar calculation). For overdoped materials we expect
at low $T$ that $\xi ^{-1}\gg T$; in this case the analysis leads to a
scattering rate proportional to $T^{2}\xi ^{2}(\theta ^{2}+\xi ^{-2})$, i.e.
one which is $T^{2}$ everywhere on the Fermi surface but is very anisotropic.

Here we have also made the additional assumption that this interaction
exists in the pairing channel, i.e. it is due to multiple scattering with
virtual Cooper pairs. These virtual Cooper pairs will lead to a
paraconductivity; we now show that this paraconductivity is negligible in
comparison to the normal carrier conductivity found in the previous section.
The paraconductivity $\sigma _{para}$ is 
\begin{equation}
\sigma _{para}=\int \frac{q^{2}d^{2}qd\omega }{(2\pi )^{3}}\frac{(\mbox{Im}%
D(\omega ,q))^{2}}{2T\mbox{sinh}^{2}(\omega /2T)}  \label{sigma_para}
\end{equation}
Evaluation and restoration of dimensional factors leads to $\sigma
_{para}\approx \sigma _{0}(T\xi )^{a}$ with $\sigma _{0}=\frac{e^{2}}{h}$
and $a=1$ if $(T\xi )\gg 1$ and $a=2$ if $(T\xi )\ll 1$; because we expect $%
(T\xi )\sim 1$ this paraconductivity is small compared to the quasiparticle
conductivity calculated previously, $\sigma \sim \frac{e^{2}}{h}v_{F}p_{F}%
\sqrt{\frac{\tau }{\Gamma _{0}}}\gg \frac{e^{2}}{h}$.

\section{Conclusion}

We have considered a model based on "cold spots"; i.e. the idea that the
quasiparticle lifetime $\tau$ is unusually short everywhere on the Fermi
line except near the zone diagonals. Direct evidence for this behavior is
found in photoemission experiments. Using a Boltzmann equation analysis we
have shown that the ansatz $\tau(\theta,T)=\frac{1}{\Gamma_0 \theta^2 +
T^2/T_0}$ with $\Gamma \approx 0.5 \;eV$ and $T_0 \approx 12\; meV$
reproduces quantitatively the observed dc and ac, longitudinal and Hall
conductivities of optimally doped materials. We regard it as particularly
significant that the model reproduces the non-Drude form of the observed
optical conductivity (both the upward curvature at low frequencies and the
roughly $1/\sqrt{\omega}$ behavior of $\mbox{Re} \sigma$ and $\mbox{Im}
\sigma$ at $\omega > 400\; cm^{-1}$) and the difference between longitudinal
and Hall scattering rates.

There are two differences between model predictions and the data. The
effects of impurity scattering are different than predicted; however a
comparison is difficult because published doping studies have involved
substitution on $Cu$ site which produces many confusing changes in the
material. Electron damage or light doping on sites away from the $CuO_{2}$
planes in {\em very clean} samples would provide a more definite test. A
more troubling discrepancy is the magnetoresistance which is predicted to
have a much larger magnitude and stronger temperature dependence than is
observed. We do not at present have a resolution, but the other successes of
the phenomenology lead us to believe one may be found. We also showed that
the anomalous lifetime could arise from the exchange of virtual Cooper pair
fluctuations. If this is the case, then the anomalous scattering would get
weaker in an applied magnetic field, perhaps reducing magnetoresistance.

Finally, we comment on the energy scale, $T_{0}$, characterizing the ''Fermi
liquid'' scattering rate $1/\tau _{FL}=T^{2}/T_{0}$. The data imply $%
T_{0}\approx 12\;meV$, a surprisingly small value corresponding to strong
scattering even along the diagonals. (Note that the ac Hall measurements of
Ref. \cite{Kaplan96} combined with the observed  $T^{2}$ Hall angle
dependence directly implies the existence of such small energy scale in the
material) Thus for $T$ of order room temperature $T^{2}/T_{0}$ is of the
order of the smallest fermion Matsubara frequency, $\pi T$, suggesting that
even along the diagonals fermi liquid behavior breaks down for $T>300\;K$.
The small value of $T_{0}$ implies that even the ''Fermi liquid'' $T^{2}$
scattering must be due to some anomalous singular scattering mechanism, for
example the virtual Cooper pair fluctuations considered in Section IV. 

{\it Acknowledgements} We thank D. Basov for sending us unpublished
data and D. Geshkenbein for helpful discussions.  A. J. M. was supported
by N.S.F. D.M.R.-9707701.


\end{multicols}


\begin{references}

\bibitem{Baskaran87}  G. Baskaran, Z. Zou and P. W. Anderson, Solid State
Comm. {\bf 63} 973 (1987).

\bibitem{Doniach90}  S. Doniach and M. Inui, Phys. Rev. {\bf B41} 6688,
(1990).

\bibitem{Emery95}  V. Emery and S. Kivelson, Nature {\bf 374} 434 (1995).

\bibitem{Coleman96} P. Coleman, A. J. Schofield and A. M. Tsvelik,
Phys. Rev. Lett. {\bf 76}, 1324 (1996).

\bibitem{Carrington92}  A. Carrington, A. P. Mackenzie, C. T. Lin, and J. R.
Cooper, Phys. Rev. Lett. {\bf 69} 2855 (1992).

\bibitem{Stojkovic96}  B Stojkovic and D. Pines, Phys. Rev. Lett. {\bf 76}
811 (1996).

\bibitem{Hlubina93}  R. Hlubina and T. M. Rice, Phys. Rev. {\bf B51}, 9253
(1995).

\bibitem{Ong97}  see e.g. Comment by N. P. Ong and P. W. Anderson, Phys.
Rev. Lett. {\bf 78} 977 (1997) and Reply by B. Stojkovic and D. Pines, ibid
978.

\bibitem{Zheleznyak97}  A. T. Zheleznyak, V. Yakovenko, H. D. Drew and I. I.
Mazin, unpublished (cond-mat/9706029)

\bibitem{Geshkenbein97}  V. B. Geshkenbein, L. B. Ioffe and A. I. Larkin,
Phys. Rev. B {\bf 55}, 3173 (1997).

\bibitem{Lee97}  P. A. Lee and X. G. Wen, Phys. Rev. Lett. {\bf 78} 4111 (1997).

\bibitem{Shen95}  Z. X. Shen and D. Dessau, Physics Reports {\bf 253}, 1
(1995).

\bibitem{Randeria97}  M. Randeria and J. C. Campuzano, unpublished
(http://xxx.lanl.gov/abs/cond-mat/9709107).

\bibitem{Andersen94}  O. K. Andersen, O. Jepsen, A. I. Liechtenstein and I.
I. Mazin, Phys. Rev. {\bf B49} 4145 (1994); see also O. K. Andersen, A. I.
Liechtenstein, O. Jepsen and F. Paulsen, J. Phys. Chem. Sol. {\bf 56} 1573
(1995).

\bibitem{Liechtenstein96}  A. I. Liechtenstein, O. Gunnarson, O. K. Anderson
and R. M. Martin, Phys. Rev. B {\bf 54,} 12505 (1996).

\bibitem{Parks95}  B. Parks, S. Spielman, J. Orenstein, D. T. Nemeth, F.
Ludwig, J. Clarke, P. Merchant and D. J. Lew, Phys. Rev. Lett. {\bf 74} 3265
(1995).

\bibitem{Geshkenbein98}  V. Geshkenbein, L. B. Ioffe and A. J. Millis,
unpublished (http://xxx.lanl.gov/abs/cond-mat/9801059).

\bibitem{Orenstein90}  J. Orenstein, G. A. Thomas, A. J. Millis, S. L.
Cooper, D. H. Rapkine, T. Timusk L. F. Schneemeyer and J. V. Waszczak, Phys. B
Rev. {\bf 42} 6342 (1990).

\bibitem{Ioffe90} L. B. Ioffe and B. G. Kotliar, Phys. Rev. {\bf B42} 10348
(1990).

\bibitem{Anderson91} P. W. Anderson, Physica C {\bf 185-189}, 11 (1991).

\bibitem{Collins89}  R. T. Collins, Z. Schlesinger, F. Holtzberg, P.
Chaudhari and C. Feild, Phys. Rev. B {\bf 39}, 6571 (1989).

\bibitem{Basov95}  D. Basov, private communication.

\bibitem{Baraduc96}  C. Baraduc, A. El Azrak and N. Bontemps, J Supercond. 
{\bf 9} 3 (1996).

\bibitem{Kaplan96}  S. G. Kaplan, S. Wu, H-T Lihn, H. D. Drew, Li Qi, D.
Fenner, J. M. Phillips and S. Y. Hou, Phys. Rev. Lett. {\bf 76} 696 (1996).

\bibitem{Clayhold97} J. Clayhold, Z. H. Zhang and A. J. Schofield, unpublished
(http://xxx.lanl.gov/abs/cond-mat/9708125). 

\bibitem{Harris95}  J. M. Harris, Y. F. Yan, N. P. Ong, P. W. Anderson, T.
Kimura and K. Kitazawa, Phys. Rev. Lett. {\bf 75} 1391 (1995).

\bibitem{Mizuhashi95}  K. Mizuhashi, K. Takenaka, Y. Fukuzumi and S. Uchida,
Phys. Rev. {\bf B52} 3884 (1995).


\bibitem{Mendels94}  P. Mendels, H. Alloul, G. Collin, N. Blanchard, J. F.
Marucco and J. Bobroff, Physica {\bf C235-240} 1595 (1994).

\bibitem{Chubukov97}  A. V. Chubukov, unpublished
(http://xxx.lanl.gov/abs/cond-mat/9709221). 

\bibitem{Ioffe93}  L. B. Ioffe, A. I. Larkin, A. Varlamov and Yu Lu, Phys.
Rev. {\bf B47} 8936 (1993).





\end{references}
\end{document}